\documentclass[preprint]{aastex}

\begin{document}
\title{HST Time-Series Photometry of the Transiting \\
Planet of HD~209458\altaffilmark{1}}
\author{Timothy M. Brown\altaffilmark{2}, David Charbonneau\altaffilmark{3,2},
Ronald L. Gilliland\altaffilmark{4}, Robert W. Noyes\altaffilmark{3},
\and Adam Burrows\altaffilmark{5}}
\altaffiltext{1}{Based on observations with the NASA/ESA 
{\it Hubble Space Telescope}, obtained at the Space
Telescope Science Institute, which is operated by the 
Association of Universities for Research in
Astronomy, Inc. under NASA contract No. NAS5-26555.}
\altaffiltext{2}{High Altitude Observatory/National Center for Atmospheric
Research, 3450 Mitchell Lane, Boulder, CO 80307.
The National Center for Atmospheric Research is sponsored
by the National Science Foundation.}
\email{timbrown@hao.ucar.edu}
\altaffiltext{3}{Harvard-Smithsonian Center for Astrophysics,
60 Garden St., Cambridge, MA 02138}
\email{dcharbonneau@cfa.harvard.edu}
\email{rnoyes@cfa.harvard.edu}
\altaffiltext{4}{Space Telescope Science Institute, 3700 San Martin Dr.,
Baltimore, MD 21218}
\email{gillil@stsci.edu}
\altaffiltext{5}{Department of Astronomy, University of Arizona, 
933 North Cherry Avenue, Tucson, AZ 85721}
\email{aburrows@as.arizona.edu}

\begin{abstract}
We have observed 4 transits of the planet of HD~209458 using the
STIS spectrograph on HST.
Summing the recorded counts over wavelength between 582 nm and 638 nm
yields a photometric time
series with 80~s time sampling and relative precision of about 
$1.1 \times 10^{-4}$ per sample.
The folded light curve can be fit within observational errors using
a model consisting of an opaque circular planet transiting a limb-darkened
stellar disk.
In this way we estimate the planetary radius $R_p = 1.347 \ \pm \
0.060 \ R_{\rm Jup}$, 
the orbital inclination $i = 86.68^\circ \ \pm \ 0.14^\circ$,
the stellar radius $R_* = 1.146 \ \pm \ 0.050 \ R_\odot$, and one parameter
describing the stellar limb darkening.
Our estimated radius is smaller than those from earlier studies, but
is consistent within measurement errors, and is also consistent with
theoretical estimates of the radii of irradiated Jupiter-like planets.
Satellites or rings orbiting the planet would, if large enough, be
apparent from distortions of the light curve or from irregularities in
the transit timings.
We find no evidence for either satellites or rings, with upper limits
on satellite radius and mass of 1.2 $R_\oplus$ and  3 $M_\oplus$,
respectively.
Opaque rings, if present, must be smaller than 1.8 planetary radii
in radial extent.
The high level of photometric precision attained in this experiment
confirms the feasibility of photometric detection of Earth-sized
planets circling
Sun-like stars.
\end{abstract}


\section{Introduction}

The low-mass companion to HD~209458 is the first extrasolar
planet found to transit the disk of its parent star \citep{cha00, hen00}.
The primary star (G0~V, $V=7.64$, $B-V=0.58$; \cite{hog00}) 
lies at at distance of 47 pc
as determined by Hipparcos \citep{per97}.
An analysis of
radial velocity measurements by \citet{maze00} gave an orbital
period of 3.524 days, with $M_{p} \sin i = 0.69 \pm 0.05 \ M_{\rm Jup}$ 
and $a = 0.0468$ AU, using the derived value of $1.1 \pm 0.1 \  M_{\sun}$ 
for the stellar mass.
When combined with the early photometric light curve data, the same
analysis yielded an orbital inclination $i = 86.1^\circ \pm 1.6^\circ$,
and a planetary radius $R_p = 1.40 \pm 0.17 \ R_{\rm Jup}$.
The planetary radius is at once the most interesting and the most
uncertain of these parameters, largely because of uncertainty in the
value of the stellar radius $R_*$.
Knowledge of $R_p$ is important because it allows inferences about
the planet's composition and evolutionary history
\citep{gui96, gui99, bur00}.
Unfortunately, the measured quantity that emerges most easily from
the photometric transit data is the ratio $R_p / R_*$, and residual
errors in the astrometry and effective stellar temperature 
suffice to make the estimate of $R_*$,
hence $R_p$, uncertain by about 10\%.
Additional small errors in $R_p$ result from uncertainties about the
stellar limb darkening.
\citet{jha00} used multicolor photometric data to reduce these
uncertainties, obtaining $R_p \ = \ 1.55 \pm 0.10 \ R_{\rm Jup}$.
Finally, analyses of Hipparcos photometric data by \cite{cas00} and 
\cite{rob00} have given refined estimates of the orbital period. 

Here we report the results of very precise photometric measurements
of transits of HD~209458~b, obtained using the STIS spectrograph on HST.
The motivations for the study were
\begin{enumerate} 
\item to obtain sufficiently accurate photometry
to reduce the ambiguity between estimates of the stellar radius, 
planetary radius, orbital inclination, and stellar limb darkening,
\item to search for evidence of planetary satellites 
or circumplanetary rings, and 
\item to search the stellar spectrum observed
in and out of transit for features imposed by transmission of starlight
through the outer parts of the planet's atmosphere \citep{sea00, bro01a}.
\end{enumerate}
We discuss the photometric results (1 \& 2) in the present work;
the spectroscopic investigation (3) will be the subject of a later paper.

\section{Observations and Data Analysis}

During a transit of HD~209458~b, the apparent brightness of the star
is reduced by a little less than 2\% as a result of the light
blocked by the gas-giant planet.
To have a useful sensitivity to smaller objects such as 
circumplanetary rings or Earth-sized satellites, 
we required photometric accuracies at least 2 orders
of magnitude better, with time resolution of a few minutes or less.
Most instruments on HST cannot meet these requirements, because
they are not designed to accept the requisite large photon fluxes.
The Space Telescope Imaging Spectrograph
(STIS), used in spectrographic mode with the CCD detector,
is the exception.
By dispersing the light in a large bandwidth over many pixels,
using the 4 e$^-$ per data number gain setting,
STIS can collect \mbox{2.5 $\times \ 10^8$ photons} per detector readout
without saturating any detector pixels.
A summation over all sampled wavelengths then provides a photometric
signal whose shot-noise-limited signal-to-noise ratio (SNR) is
above $10^4$.
The readout time for a detector subarray of the necessary size
($1024 \ \times \ 64$ pixels) is about 20 s; it is thus possible
to maintain both an 80 s sampling cadence and a respectable duty
cycle of 75\%.
To attain near shot-noise limited operation, one must control, or
verify the smallness of, numerous sources of instrumental noise,
such as variations in the shutter opening and closing times, and 
CCD gain variations combined with spectrum motion on the detector.
One of us had already investigated the use of the
STIS as a precise photometer, and had demonstrated that, with care,
these systematic errors can be suppressed \citep{gil99a, gil99b}.
We note that these studies further demonstrated that time series with
SNR = $10^4$
could be maintained even well past saturation of individual pixels,
although in the present observations we did not saturate the detector.

We observed HD~209458 during each of 4 planetary transits, on
UT dates 2000 April~25, April~28/29, May~5/6, and May~12/13.
The full duration of a transit is 184.25~minutes,
or slightly less than 2 HST orbits, which are 96.5~minutes each.
Objects orbiting the planet may, however,
precede or follow it by as much
as $R_H$, the radius of the planet's Hill sphere
(i.e., the radius at which the star's tidal forces would dislodge a satellite):
\begin{equation}
R_H \ = \ a \left ( M_{p} \over 3 M_{*} \right ) ^{1/3}
\ = \ 4.2 \times 10^5 \ {\rm km} \ = \ 5.9 \ R_{\rm Jup} \ \ ,
\end{equation}
where $a$ denotes the planet's orbital radius and
$M_{p}$ and $M_*$ the masses of the planet and star.
Since the planet's orbital speed is estimated to be 143~km~s$^{-1}$,
one must search for gravitationally-bound objects as much
as 49 minutes, or about one half of an HST orbit,
before and after the transit central time.
This requires 4 full HST orbits to assure adequate coverage
before and after the planetary transit.
To allow the telescope pointing and thermal environment to
stabilize before beginning critical observations, we added
one further orbit at the beginning of each transit sequence,
for a total of 5 HST orbits for each transit.
The orbital phasing was such that each observed transit contained
2 initial orbits that were completely off-transit, 2 orbits that fell
almost entirely during the transit, and one following
orbit that was off-transit.

We took all observations using the G750M grating, covering
the wavelength range $581.3 \le \lambda \le 638.2$ nm,
with a resolution of $R = \lambda / \Delta \lambda = 5540$,
corresponding to a resolution element of 2.0~pixels or 0.11~nm.
This wavelength range was chosen to cover the region of the
Na~D lines, where a large signature from transmission through
the planetary atmosphere is thought to be likely.
We used a $52 \ \times \ 2$ arcsec slit, much larger than
the (typically 0.07 arcsec full width at half maximum) 
stellar image size, to minimize
variations in the fraction of stellar light lost at the slit
edges due to minor guiding and focus changes.
(We did not use the ``clear'' slit in order 
to avoid excessive sky background.)
The CCD area read out consisted of a 1024 $\times$ 64 subarray,
covering the entire possible range in the dispersion direction
and about 3.2 arcsec across the dispersion.
Exposure times were 60~s, which with a 20~s readout time gave
an observing cadence of 1 sample per 80~s.
This exposure time gave about $1.55 \times 10^8$ detected
photons per spectrum, which would correspond to an optimal
photon-noise-limited precision of $8.0 \times 10^{-5}$,
or $87 \ \mu$mag.
Figure 1 shows a typical extracted stellar spectrum (see the discussion
below for details of the extraction process).
The first orbit of each group of 5 has a reduced number
of spectra due to time spent on target acquisition.
During each visit of five orbits, we obtained 28 spectra
during the first visit, then 36 spectra during each of the
subsequent three orbits, then 35 spectra during the final orbit,
for a total of 171 spectra.  The final data set thus 
contained 4 $\times$ 171 = 684 individual spectra of HD~209458.
We took wavelength calibration spectra just before each of the
5 orbits, and, as an associated calibration program,
we obtained 49 flat-field exposures during occultation
time within the orbits surrounding the transit observations.
All data for both the science program (8789) and 
calibration program (8797) are publically available via the
HST archive\footnote{See http://archive.stsci.edu.}; there is
no proprietary period for these data since they were obtained
through Director's Discretionary time.

Observations of the first transit (UT 2000 April 25) were partly 
compromised by a database error in the location of the subarray
for this rarely-used secondary central wavelength setting of G750M. 
The result was that the spectrum was not
entirely contained within the CCD subarray for the red half
of the spectral range.  This error led to reduced counts and increased 
sensitivity to cross-dispersion positional drifts over this range.
In the subsequent data analysis, we ignored this part of the wavelength
range, but only for the first transit.
Rapid work by the staff at STScI identified the source of this
error and provided an effective correction strategy (a real-time
slew executed after target acquisition for the visit on April 28/28,
and a correction to the database thereafter), so that the
spectra for subsequent transits were correctly positioned on the
subarray.

Data reduction consisted of
\begin{enumerate}
\item recalibrating the 2-dimensional CCD images,
\item removing cosmic ray events,
\item extracting 1-dimensional spectra, 
\item summing the detected counts over wavelength to yield a photometric index 
\item correcting the resulting photometric time series for 
variations that depend on the phase of the HST orbit, and
for variations between visits
\end{enumerate}

Recalibration of the CCD images started from the bias-
and dark-subtracted
and flat-fielded images
produced by the standard STIS data reduction pipeline.
The STIS pipeline builds a weekly bias image from exposures
taken for this purpose.  We examined the biases that 
were subtracted from each image and found them
to be sufficiently accurate.
In contrast to this, the pixel-to-pixel flat field images are
often produced from data that are several months old, and
we found that the relative sensitivity of numerous pixels
had changed since the most recent pipeline flats were
produced.  Furthermore, the pipeline flats did not have a sufficiently
high SNR for our purposes.  We requested that high SNR flats
be taken during times of Earth-occultation contemporaneous
with our science exposures.  We shot two sets of 7 such exposures 
for each of the visits on UT 2000 April 24, 27, \& May 12,
and one set of 7 such exposures for the visit on UT 2000 May 5.
For each set of 7 exposures, the intensity in each exposure
increased with time and in a reproducible manner from one
set to the next.  
The origin of this increase is unknown, although a possibility is
that the calibration lamp warmed slowly as the series of exposures progressed.
This effect was removed by renormalizing
by the total number of counts.  The relative illumination
among pixels by the continuum source was remarkably constant:
No temporal drift was observed (aside from the normalization
variation described above), thus we were able to quantify the
relative sensitivity of individual pixels (which typically
varied with an amplitude of 1\%) to a precision of better than
than 0.1\%.  Between each visit, a small fraction of the
pixels changed their relative sensitivity significantly, most likely
due to radiation events between visits.  Furthermore, the STIS
CCD was annealed on UT 2000 May 6, between the third and fourth
visits, which changed the behavior of a number of pixels.
To handle sensitivity variations between any two visits, 
we tagged all pixels whose sensitivity we observed (from the
contemporaneous flats) to have changed, and produced new flat 
fields for each visit where these
pixels were replaced by their value derived from the 
data obtained during only the one visit.  We then multiplied all the
science exposures by the pipeline flat field, to undo
the calibration from the pipeline, and then divided all
the exposures by our newly derived flat fields.
The calibration changes we imposed were typically small, typically 
a fraction of a percent in the gain of individual pixels.  
Since the photometric signal (described
below) is produced by summing over a large number of pixels,
these changes made only a small, but noticeable, improvement 
in the final results.
In contrast to this, the improved precision from this recalibration
provided significant gains in the analysis of the spectra to search
for transmission features (to be presented in a separate paper).

Cosmic ray hits on the CCD were a significant source of noise in
the recalibrated data.
We corrected for these by forming a time series of the intensity at
each pixel and fitting this series as a fourth-order polynomial
in time.
Points showing statistically significant positive differences from
the fit were deemed to be corrupted.
Such points were replaced by the
value of the fitted function at the corresponding time.
Of the 65536 pixels in each image, typically 36 (this is the median number) 
had to be corrected in this way, and of these pixels about 
25\% contributed to the final photometric index for the image.

We extracted 1-dimensional spectra simply by summing along
CCD columns, taking at each column
a band of fixed width centered on the measured
cross-dispersion spectrum position.
Since motion of the spectrum in the cross-dispersion
direction was measured to be much less than one pixel (typically 
motions were roughly 0.05~pixels within visits) for the entire data set,
we used the same position for the integration band to produce 
all extracted spectra. 
We used a total cross-dispersion band 17~pixels in
width, which we found produced the minimum rms variation
in the photometric time-series for visits that occurred out of transit.
The result of this operation was a 1-dimensional spectrum
sampled at 1024 points.

%
For the most part
we describe here a single photometric index, namely a sum
$S(t)$ of the 1-dimensional spectrum over almost the entire available
spectral range (581.9 to 614.6~nm for the first transit, and
581.9 to 637.6~nm for the others).
The only refinement in this process was to position the endpoints
of the summation so as to avoid obvious absorption lines,
so that the sum would be insensitive to displacements along the dispersion.
In addition
we performed an experiment in which we summed separately over the red and blue
halves of the wavelength range; the difference between these two light
curves results from the color dependence of the stellar limb darkening.
The details of this analysis are described in section 3.3.

The time series $S(t)$ shows small but repeatable variations
in phase with the HST orbital period.
We do not know the origin of these variations, though we
suspect that they are connected with the telescope's orbital
thermal cycle.
The variations are fairly well approximated by a linear
decrease with an amplitude of 0.1 \% over 48 minutes, with some curvature
at the beginning and end of each orbit.  
We corrected the time series on a transit-by-transit basis.
For each transit, we phased the data from the out-of-transit
orbits (orbits 2 and 5) to the HST orbital period.
We then fit a fourth order polynomial to these data,
and divided all five orbits of the transit by this function.
The first sample in every orbit is always smaller than the 
average by 0.25\%; we rejected these values from the time series.

The corrected time series have transit-to-transit differences
of scale that are large for the first transit (because we
summed over a smaller range of wavelength) and smaller (typically
0.1\%) for the remaining 3 transits.
Lacking a comparison star or other external calibration source,
we cannot say how much of this long-term variation arises
from the star and how much from instrumental drift.
For our purposes this is not important, however, since we are
primarily concerned with variations on time scales of minutes
or hours, and not weeks.
These scale variations were also removed by the
procedure described in the previous paragraph.

The normalized time series have a value of unity when
averaged over the out-of-transit orbits, and
minimum values near the transit centers of 0.9836.
Figure 2 shows the time series for each of the 4 transits
phased with respect to the planetary orbit,
and Figure 3 shows the combined data for all 4 phased transits,
excluding the first orbit of each visit.
One may estimate the noise in Figures 2 and 3 from the scatter
among the out-of-transit observations.
For the last 3 transits, these are typically $1.1 \times 10^{-4}$,
or 120~$\mu$mag for each 60~s integration;
the noise for the first transit is worse by a factor of
about 2.3, mostly because of increased sensitivity to transverse
motions of the poorly-centered spectra.
This precision suffices to show the 1.64\% transit
dip with a SNR of about 150, at high time resolution.

\section{Interpretation}

\subsection{Planetary Orbital Parameters}

Ideally, we would like to measure $T_{c}$, the time at
the center of the transit, for each of the 
four observed transit events.  However, we have gaps
in our observations of any one transit (see Figure 2)
due to Earth occultation that prevent us from doing so
with sufficient accuracy.
We can, however, estimate the period accurately (and independently
from the absolute times of the transits) by seeing which
assumed value of the period produces the minimum scatter
in the phased transit curve.  In this manner we derive an orbital
period of $P = 3.52474 \pm 0.00007$~d.  Having evaluated
the period, we can then subtract the appropriate multiple
of it from each of visits 2, 3, \& 4 (1, 3, \& 5 planetary orbital periods
later, respectively), to phase the data to the time
of the first transit.  We then fold the transit curve 
about an assumed time $T_{c}$ for the midpoint,
and derive $T_{c}$ from minimizing the scatter between
the observation before and after the center of transit.
We derive $T_{c} = 2451659.93675 \pm 0.0001$~HJD.
We can then compare this value of $T_{c}$ with the three
other accurately measured transit times, two from \citet{cha00} 
and one from \citet{jha00}.  Doing so, we derive a more accurate 
value of the period, due to the large number of transits
that have now elapsed from these observation from fall 1999.
We find $P = 3.52480 \pm 0.00004$~d.  The uncertainty in
the period is 4.3~seconds.  Our derived value is in excellent
agreement with the values derived with similar precision 
from Hipparcos archive photometry
by \citet{cas00}, who found $P = 3.524736 \pm 0.000045$~d, and 
by \citet{rob00}, who found $P = 3.524739 \pm 0.000014$~d.

\subsection{Stellar and Planetary Parameters}

To estimate the planet's radius, we sought to represent
the light curve as the result of an opaque sphere of
radius $R_{p}$ in an inclined circular orbit about a
limb-darkened star of radius $R_{*}$.
We took the stellar mass to be $1.1 \pm 0.1 \ M_{\sun}$ \citep{maze00}, 
with the orbital radius and velocity determined from the period
and Kepler's laws.
Free parameters in the fit were $R_{p}$, $R_{*}$, the orbital
inclination $i$, and the parameters $u_{1}$ and $u_{2}$ 
describing the stellar limb darkening in the form
\begin{equation}
{I(\mu) \over I(1)} = 1 - u_{1} (1 - \mu) - u_{2} (1 - {\mu})^{2}  \ \ ,
\end{equation}
where $\mu$ is the cosine of the angle between the
line of sight and the normal to the local stellar surface \citep{cla90}.
A simpler and often-used parameterization replaces the right hand side 
of Eq. (2) with
$1 - u (1 - \mu)$.
We find that this parameterization worsens the $\chi^2$ statistic
of the fit slightly, but significantly.
We note, however, that fitting the expression in Eq. (2) gives
strongly anticorrelated formal errors for $u_1$ and $u_2$.
Thus, a more natural set of limb darkening parameters is
$(u_1 + u_2)$ and $(u_1 - u_2)$.
The former describes the total magnitude of the limb darkening
variation, and is well constrained by the observations, while
the latter describes the amount of curvature in the limb darkening
function, and is relatively poorly constrained.
For explanatory purposes below, we shall refer to these combinations
rather than to $u_1$ and $u_2$ individually,
and moreover we shall adopt the approximation that $u \simeq u_1 + u_2$.

The conceptual basis for this fitting process is illustrated in
Figure 4.
At the lowest level of approximation, the light curve is
described by just 2 parameters: its depth $d$ depends mostly
upon $R_p/R_*$, while its duration $l$ depends mostly upon
the transit's chord length, and therefore upon $R_*$ and $i$.
At this treatment level there are fewer observables $\{d,l\}$ than
unknowns $\{R_{p},R_{*},i\}$, and so one must estimate $R_*$ from other
evidence in order to obtain values for $R_p$ and $i$.
This fitting degeneracy can be removed by taking account of
more subtle effects:
The duration $w$ of the planet's ingress and egress depends
upon $R_p$ but is also proportional to $\sec \psi$, where $\psi$
is the angle between the planet's line of motion and the
local normal to the stellar limb.
Thus, $w$ depends upon $R_{p}$, $R_{*}$, and $i$.
Finally, the curvature $C$ of the light curve between second
and third contacts depends upon the stellar limb darkening
parameter $u$ and upon $i$ and $R_{*}$.
Thus, if $d$, $l$, $w$, and $C$ can be measured with adequate
precision, one may estimate each of the 4 independent system
parameters $R_*$, $R_p$, $i$, and $u$.
Fitting for both $u_1$ and $u_2$ requires, in addition to the
quantities already mentioned, a measurement of the detailed shape
of the light curve between second and third contacts.
All of the foregoing assumes that the star's mass $M_{*}$, 
and hence the planet's orbital velocity and semi-major axis, 
are known.  The derived value of $R_{p}$ is only weakly dependent
upon the assumed value of $M_{*}$, scaling as 
$R_{p} \propto {M_{*}}^{1/3}$. 
We assume $M_* \ = \ 1.1 \pm 0.1 \ M_{\sun}$ \citep{maze00},
and our derived errors include this uncertainty in the stellar 
mass.  

We derive best-fit values for $\{R_{p}, R_{*}, i, u_1, u_2\}$
by minimizing the ${\chi}^{2}$ of the fit.  The reduced ${\chi}^{2}$
for the best fit values was 1.07, indicating that the
model is a good fit to these data.
To derive 1$\sigma$ errors for each parameter, we change the 
value of that parameter and fix it at a new value, 
and then allow all other parameters to float, as well
as allow for a stellar mass between 1.0 and 1.2 $M_{\sun}$.
We repeat this procedure until the best fit solution
produces an increase in the ${\chi}^{2}$ corresponding to
a 1$\sigma$ change.  The best-fit parameters and their 
errors are given in Table 1.
Both the stellar and planetary radii are found to be
in agreement with those derived by \citet{maze00},
and the precision has been increased greatly.
Most significantly, the error in the planetary radius
has been reduced from $0.17 \ R_{\rm Jup}$ to
$0.06 \ R_{\rm Jup}$.  We note also that our derived 
value for $R_{*}$ is in agreement
with that derived from the {\it Hipparcos} distance, 
$R_{*} = 1.18 \pm 0.09 R_{\sun}$ (see Jha et al. 2000 for
details).
The values for $R_{p}$ and $R_{*}$ derived here are
somewhat smaller than those derived by analysis
of multicolor data by \citet{jha00}, although the differences 
are consistent with the errors, differing by less than 1.4$\sigma$.
There are some correlations among the formal errors for the
various derived parameters, and these are illustrated in Figure 5.
We reiterate that our stated 1$\sigma$ errors include, and
are dominated by, these correlations.  
In particular, the uncertainties in the radii of the planet and star
are dominated by the assumed uncertainty of $\pm 0.1 \ M_{\sun}$
in the stellar mass.

\subsection{Color Dependence of the Light Curve}

To check for color dependence of the transit curve,
we divided the spectral region in two ranges,
581.9--598.3~nm (``blue''), and 598.3--637.6~nm (``red'').
We then generated photometric timeseries for each
of these following the
procedure described in Section 2.  We used data from 
only visits 2, 3, and 4, since the data from 
visit 1 lacked the red half.  For each color,
we phased the data to the period of 3.52474~d, and 
folded it about the midpoint of the transit, $T_{c}$.  
Since we do not have useful data from the first visit,
gaps exist in the time coverage.
We then grouped the data into 5-minute bins, and 
for the data in each bin, we fit brightness as a linear function of time
within the bin.
We took the brightness for the bin to be this linear function,
evaluated at the 
centeral time for the bin.
We then differenced the red and blue transit curves
generated in this manner.  These data are
shown in Figure 6.  From these data, it is clear that
the transit is deeper in the blue at times near the
center of transit, and deeper in the red at times
when the planet is near the limb, as would be expected
from the greater limb-darkening in the blue (see, e.g.,
\citet{ros71, sac99} for sample 
differenced-color light curves).  

To generate the best-fit 
light curve for the difference of these color data,
we fixed $R_{p}$, $R_{*}$, and $i$ at the values
derived in Section 3.1.  We then allowed small changes
to the parameters describing the limb darkening, 
such that for the red half (denoted by $R$), 
\begin{eqnarray}
{(u_{1} + u_{2})}_{R} = (u_{1} + u_{2}) + {\alpha} \nonumber \\
{(u_{1} - u_{2})}_{R} = (u_{1} - u_{2}) + {\beta},
\end{eqnarray}
and for the blue half (denoted by $B$),
\begin{eqnarray}
{(u_{1} + u_{2})}_{B} = (u_{1} + u_{2}) - {\alpha} \nonumber \\
{(u_{1} - u_{2})}_{B} = (u_{1} - u_{2}) - {\beta}.
\end{eqnarray}
We then generated model light curves for both the
red and blue data, differenced these, and
evaluated the ${\chi}^{2}$ of this fit 
to the data.  The data are best fit
by ${\alpha} = -0.021$ and ${\beta} = 0.063$,
thus the derived values for the limb-darkening are
\begin{eqnarray}
{(u_{1} + u_{2})}_{R} = 0.619 \pm 0.03 \nonumber \\
{(u_{1} + u_{2})}_{B} = 0.661 \pm 0.03 \nonumber \\
{(u_{1} - u_{2})}_{R} = 0.003 \pm 0.1 \nonumber \\
{(u_{1} - u_{2})}_{B} = -0.123 \pm 0.1, 
\end{eqnarray}
consistent with the measured Solar limb-darkening \citep{cox99}.
We show this best-fit curve in Figure~6.  
Assuming the best-fit values for $i$ and $R_{*}$,
we can calculate the projected separation of the
center of the planet and the center of the star
at any time during the transit, and thus at any
point in this color curve.  This axis is shown
at the top of Figure 6.  From this one can determine the 
fractional stellar radius at which the limb-darkening curves
(weighted over the stellar disk) cross; this occurs near 
a radius of $0.84~R_{*}$.

The color-dependent amplitude shown in Figure~6
is small primarily because we are contrasting the 
limb-darkening over two bands that are separated by
only 28~nm.  Observing the transit in very disparate
bands across the visible and near-IR would show
a much greater effect.  If one were willing to assume
a model for the limb-darkening of the star,
one could use this effect to break the degeneracy
between the parameters $R_{p}$, $R_{*}$, and $i$,
even if the data did not have the precision of
that which we present here.  This is precisely
what was done with multicolor observations of HD~209458
in Jha et al. (2000).

\subsection{Search for Circumplanetary Rings}

If the planet of HD~209458 were circled by a
ring system with significant opacity, 
the rings would cause distortions of the light
curve relative to that of a spherical body \citep{sch99}.
The cross-sectional area of the planet would appear larger
because of the light obstructed by the rings,
and (more usefully, for detection purposes) 
one would also see small dips in the light
curve before first and after fourth contact (see Figure 7).
The phased light curve is fit within observational errors by
the simple planetary-transit model described above,
so there is no evidence for rings in the current data.
We can, however, set an upper limit on the size of a ring system
consistent with the observations.
For this purpose we assume that such a ring system lies
in the planet's orbital plane, that it extends continuously from the planet's
limb to a maximum radius $R_r$ (measured in units of $R_{p}$),
and that it is entirely opaque to transmitted light.
We cannot simply assume the best-fit values for $\{R_{p},R_{*},
i,u_1,u_2\}$ from section 3.1,
because, if circumplanetary 
rings are present, some area is occulted by the rings;
this would cause us to over-estimate $R_{p}$.
To simplify the investigation, we fix $R_{*}$, $u_1$, and $u_2$.  Changing
the value of $R_{*}$ does not affect the results since it
would result in a larger value of $R_{p}$, and we state 
an upper limit for $R_{r}$ in units of $R_{p}$. 
For each trial value of $R_{r}$, we allow $R_{p}$
and $i$ to float, and derive an upper limit for $R_{r}$
by finding the value above which the ${\chi}^{2}$ increases
by an amount corresponding to a 3$\sigma$ change.
The maximum ring radius consistent with the observations at this
confidence level is then
1.8~$R_{p}$;
this is slightly smaller than the radius
of Saturn's ring system, measured in units of Saturn's radius.
This fairly low sensitivity to ring systems results mostly
from the assumption that rings must lie in the planet's orbital
plane, and hence that the ring plane must be nearly edge-on as seen
from Earth.

\subsection{Search for Planetary Satellites}

A satellite orbiting HD~209458~b might be detectable either from its
photometric signature or from its influence on the orbital motion
of the planet itself \citep{sart99}.
A satellite would block light in addition to that obstructed by the
planet, unless it happened to be projected onto the planet's disk
during the transit.
This additional obstruction could occur either earlier or later than
the main transit, depending upon the satellite's position in its
orbit.
Similarly, its duration could be slightly different from that of
the planetary transit, because the satellite can move significantly
in its orbit during the transit, and because the chord the
satellite strikes across the star may be 
longer or shorter than that of the planet.

As was the case for rings, there is no evidence for satellites in
the photometric time series.
Figure 8 shows the residuals about the best-fit light curve for each
of the observed transits, along with transit curves that might be
expected from a satellite with 1.5 times the radius of the Earth
and an orbital period of 1.5 d.
A signal of this size would be easily detected, if it were present.
To set a better limit on the size of possible satellites, we searched
the residual data for repeated transit-like events by applying
matched filters that simulated the light curves from satellites
with a range of orbital periods, phases, and semi-major axes.
This process was analogous to that used to search for transiting
planets in observations of
the eclipsing binary CM Dra \citep{doy00}, and in
HST time-series photometry of the globular cluster 47 Tucanae
\citep{gil00, bro01}.
The process we actually used worked as follows: 
We first subtracted the best-fit model for the 
transit curve of the planet from the data, yielding
the photometric residuals $R(i)$, with $i = [1,...,N]$.
We used the data only from the last three visits, 
as the errors due to the offset
of the spectrum in the CCD subarray during the first transit
are relatively very large.  We constructed a dense grid of
sample satellite orbital periods ($p \in [ 1~d, \ 3.5~d ]$) 
and satellite orbital phases (${\phi} \in [ 0, 1 ]$). 
For each pair of $\{ p, {\phi} \}$, we evaluated, for
each visit, the amount of time by which the center of the 
satellite transit would lead or trail the center of the 
planetary transit.  We then subtracted these corrections
from the times of the observations (so as to 
phase the data to any potential satellite transit),
yielding the new times $t_{p,{\phi}}(i)$.  Neglecting small
changes in the duration of the satellite transit due
to satellite orbital motion during the transit times,
we assumed the duration of the satellite transit, $d_{sat}$,
to be the same as that of the planetary transit, 184.25~minutes.
We then computed the correlation of these data with a box-car
function defined by:
\begin{displaymath}
B_{p,{\phi}}(i) = \left\{ \begin{array}{ll}
1 & \textrm{if $|t_{p,{\phi}}(i)| \le d_{sat}/2$}\\
-1 & \textrm{if $|t_{p,{\phi}}(i)| > d_{sat}/2$}
\end{array} \right\} .
\end{displaymath}
The correlation was given by
\begin{equation}
C_{p, {\phi}} = \frac{2 \sum_{\rm i=1}^{N} B_{p,{\phi}}(i) R(i)}{N},
\end{equation}
where the leading factor of 2 accounts for the fact that
any such satellite transit will have a mean of zero,
since we had already subtracted the best-fit planetary
transit.  In the case of noiseless data, 
the value of $C_{p,{\phi}}$ at the correct $\{ p,{\phi} \}$ 
is the depth of the transit:  For a satellite of radius $R_{sat}$
passing in front of HD~209458, we would have 
$C_{p,{\phi}} \simeq 6.4 \times 10^{-5} (R_{sat} / R_{\oplus})^{2}$.

To derive detection thresholds,
we replaced all the residuals with numbers drawn from
a Gaussian distribution with ${\sigma} = 1.14 \times 10^{-4}$
and evaluated $C_{p,{\phi}}$.  We repeated this many times,
and evaluated the maximum value of $C_{p,{\phi}}$ for each 
fake data set.  
These maximum values were Gaussian-distributed, with the upper 3-$\sigma$
point of the distribution lying at $C=5.2 \times 10^{-5}$.
Thus, assuming the noise to be purely Gaussian, we exclude with
99.7\% confidence the presence of satellites larger than 
the corresponding radius, namely $R_{sat} \ = \ 0.9~R_{\oplus}$.

The above conclusion is correct only if the errors are indeed
Gaussian.  However, we can see from the residuals
that there may exist orbit-to-orbit drifts.  To account
for these as well, we performed a different test.
For each orbit, we calculated the average of the residuals,
and found the maximum average offset to be $6.9 \times 10^{-5}$.
As before, we replaced all the residuals with numbers drawn from
a Gaussian distribution with ${\sigma} = 1.14 \times 10^{-4}$.
We then added to each orbit of data an additional offset
drawn from a uniform distribution between $\pm 6.9 \times 10^{-5}$.
We then evaluated $C_{p,{\phi}}$, and repeated this
procedure many times.  Following the same procedure as before,
we found a $3{\sigma}$ limit of $C=9.0 \times 10^{-5}$,
corresponding to $R_{sat} = 1.2~R_{\oplus}$.
We believe that this noise model better describes our data than does
pure Gaussian noise, so we prefer this estimate of the detection threshold.
The smaller threshold remains of interest, however, as an indication
of the size of detectable companions, should the remaining systematic 
errors of measurement
be eliminated.

For our actual residuals, the maximum of $C_{p,{\phi}}$ 
is $4.88 \times 10^{-5}$.
This value is significantly 
below the detection threshold described in the last paragraph,
and consistent with the result of pure noise.  We conclude
that we have no evidence for a satellite orbiting
the planet of HD~209458, and can exclude (at the $3{\sigma}$ level)
satellites larger in radius than $1.2~R_{\oplus}$.
These results show that if we obtained the same precision
on a star with $R_{*} = 1.0~R_{\sun}$ (recall that for HD~209458,
$R_{*} = 1.15~R_{\sun}$), we would be able to make a 3$\sigma$ 
detection of a $1.0-R_{\oplus}$-sized planet in
transit across its star.
In the absence of orbit-to-orbit drifts, the limit would be $0.78~R_\oplus$.

Satellites might also be detected by the periodic displacement of
the planet in its orbit due to the gravitational attraction of
the satellite.
The magnitude $\delta x$ of this displacement could be as much as
$\delta x \ = \ a_{sat} M_{sat}/M_{p}$, where $a_{sat}$ is the satellite's
orbital radius and $M_{sat}$ is its mass.
The visible effect would be to wobble the planet ahead of or behind
its mean orbital phase, assuming the satellite orbit to be
approximately coplanar with that of the planet.  
Transits would therefore occur early or
late relative to the ephemeris, depending upon the phase of the
satellite in its orbit.
For an Earth-mass satellite orbiting HD~209458~b at a distance of
one Hill sphere radius, the maximum temporal excursion is 13 s.
This time is comparable to the formal 3$\sigma$ error on the
estimate of the central time of a single transit.
Relative to the time predicted from a best-fit estimate
(based on these HST data only) of
the period and initial epoch,
the observed timing displacements (seconds) for the 4 transits are
$\{-20.0 \pm 10.0, \ 0.3 \pm 4.4, \ -3.3 \pm 6.2, \ 10.0 \pm 4.6 \}$.
Figure 9 shows these observed displacements.
Also shown is the ${\chi}^{2}$ statistic for each transit, as a
function of displacement from the observed transit times.
From these curves, it is evident that
the differences between the observed transit times and the
predicted ones are at most about 2$\sigma$.
We doubt these discrepancies arise from satellites of the planet.
Rather, we suspect that small systematic errors in the
observations (for example, small linear trends in detector sensitivity
that run from beginning to
end of a 5-orbit transit observation) may be responsible.
The residuals shown in Fig. 8 support this conclusion, with
the residuals for the transit on 12 May being predominantly negative
before the transit and positive after it; residuals due
to a planetary satellite should affect the in-transit data
(orbits 3 and 4), but not the out-of-transit data (orbits 2 and 5).
Taking the observed transit time variations to be an upper limit
on the displacement caused by an unseen satellite,
and assuming its orbital radius to be that of the Hill sphere,
we can exclude with 3$\sigma$ confidence the presence of satellites of more
than 3 Earth masses.

\section{Discussion}

The planetary radius 1.347~$R_{\rm Jup}$ inferred from the 
HST light curve is consistent with a previously reported value
of 1.40~$R_{\rm Jup}$ \citep{maze00, hen00}, and somewhat smaller
than the value of 1.55~$R_{Jup}$ derived from multi-color data
by \citet{jha00}.
The new estimate is more likely 
to be correct, because the radius of the parent star is determined
as part of the fitting process, rather than being assumed.
Indeed, the works just cited identified uncertainty in the
stellar radius as an important contributor to the uncertainty
of the final result.
An assumption about the stellar mass is required in any case, 
but our errors include a 10\% uncertainty in the
stellar mass.  Furthermore, the derived value for the 
planetary radius is only weakly dependent upon the value
of the stellar mass.
A radius of 1.35~$R_{\rm Jup}$ is consistent with the irradiated
model described by \citet{bur00};
it falls between the models with high (0.5) and low (0.0) Bond 
albedo, and the error of $0.05 \ R_{\rm Jup}$ is, in principle,
small enough to distinguish between these two albedos.
Implied values of the mean density $\rho$, surface gravity $g$, and escape
velocity $v_{e}$ all increase relative to earlier estimates, but not
by large amounts.  
We find $\rho = 0.35 \ {\rm g \, cm^{-3}}$, $g = 943 \ {\rm cm \, s^{-2}}$, 
and $v_{e} = 43 \ {\rm km \, s^{-1}}$.
These changes are all such as to increase the estimated stability of the
planet against disruption by tidal forces, thermal evaporation, or mass
stripping by the stellar wind.

Although it was possible a priori for the system to have observably
large satellites or rings, the absence of these features is not surprising.
Only large satellites (bigger than the Earth) could have been
detected by these observations,
and for a satellite to survive so close to the star, it would
have to be made of refractory materials.
The solar system contains no bodies that meet both of these
requirements (Earth and Venus come the closest),
so perhaps it is reasonable to guess that HD~209458~b likewise is not
home to such an object.
Similar comments apply to a ring system, which would have to
be large and opaque in order be detectable in our observations.
In the harsh radiation environment 0.05~AU from the central star,
the processes that destroy rings of fine particles would be accelerated,
and long lifetimes may not be expected.
One must remember, however, that the present observations could not
have detected the Galilean satellites of Jupiter, and
(because of their small optical depth) the rings of Saturn would
be only marginally visible.
Our observations constrain the presence of such objects only to
the extent that large and obvious companions are excluded.

In addition to their direct scientific interest, these observations
provide the best example to date of the capabilities of extremely precise
photometry from space.  This suggest a number of potentially
rewarding future observations:

1. With the achieved precision, it may be possible to detect the
reflected light from a close-in giant planet \citep{sea98, sud00, mar99}.  
To date, such studies \citep{cha99, col99} have produced 
only upper limits,
and have required restrictive  assumptions about the
spectrum of the reflected light.  The amplitude of
this effect in the HD~209458 system would be 
${\epsilon}_{\lambda} = p_{\lambda} (R_{p}/a)^{2} =  
p_{\lambda} \times 2.0 \times 10^{-4}$, where $p_{\lambda}$ is the 
wavelength-dependent geometric albedo.  The transiting configuration
of HD~209458~b would make it the ideal target, since at times
of secondary eclipse (ie. just as the planet passes behind
the star), the planet would pass within 26 minutes
from being nearly fully illuminated
to not visible.  For such observations where there is
no particular interest in moderate resolution spectroscopy
(as was the case for this program), it would be desirable
to switch to the low resolution gratings, and thus measure
the albedo across the wavelength range where the central
star outputs the majority of its energy.  One could thus
evaluate the net energy deposition into the planet, a key
quantity in understanding its evolution \citep{bur00}.

2. The perturbation caused by additional planets in the HD~209458 system
would change the observed times of transit.  A 1~$M_{\rm Jup}$
planet at 10~AU would cause the central star to move
$\pm$~0.01~AU, and thus the transits would be observed
as much as 5~s earlier or later.  This effect could be used
to infer the presence of such additional companions,
although many years would be required to observe the
effect, due to the long orbital periods at these
large semi-major axes.

3. Although the STIS instrument was not designed with high SNR photometry in
mind, the achieved precision of the photometric time series
confirms that it is feasible to detect the transits of Earth-sized
planets across the disks of Sun-like stars.
Moreover, reaching this precision does not depend upon the HST's
large aperture;
the limited bandwidth used in this experiment and (to a lesser degree)
the transmission losses in the spectrograph combined to make the
photometry far less efficient than it could be.
For example, with a bandwidth spanning 400--1000~nm and a system
efficiency of 50\%, the photon count rates achieved here could be
reached with a telescope of only 25~cm aperture.
Placed in a suitable orbit, so that full uninterrupted transits
of HD~209458~b could be observed,
such a telescope could detect a satellite of the mass
or radius of Ganymede after observing about 100 transits,
which is to say, within about 1 year.
The first generation of orbiting telescopes designed for such
purposes are now under development
(MOST, \citet{mat00}; COROT, \citet{mic00}; MONS, \citet{kje00}),
and larger-scale projects are planned or proposed, such as The Kepler Mission, 
\citep{koc98}, 
and Eddington\footnote{See http://astro.esa.int/SA-general/Projects/Eddington/.}.

With the new photometric and spectroscopic data sets that will
soon become available, we can look forward to an exciting decade
unraveling the structure and history of the close-in extrasolar planets.

\acknowledgements
We are grateful to the HST STIS and operations teams,
especially Helen Hart, Gerard Kriss, and Jeff Valenti
for their prompt and insightful help in resolving the
database error experienced during the first transit.
Support for proposal \#HST-GO-08789.01-A was provided by NASA through
a grant from the Space Telescope Science Institute, which is operated
by the Association of Universities for Research in Astronomy, Inc.,
under NASA contract NAS5-26555.
Further support for this work was provided through NASA grants
NAG5-7073 and NAG5-7499.
D.~C. is supported in part by a Newkirk Fellowship of the High Altitude
Observatory.

\clearpage

\begin{deluxetable}{lc}
\tablecaption{HD~209458~b Fit Parameters}
\tablewidth{0pt}
\tablehead{\colhead{         } & \colhead {                            }}
\startdata
$R_*$ &1.146 $\pm$ 0.050 $R_{\odot}$\\
$R_p$ &1.347 $\pm$ 0.060 $R_{\rm Jup}$\\
$i$ &$86.68^{\circ}$ $\pm$ 0.14$^\circ$\\
$(u_1+u_2)$ &0.640 $\pm$ 0.030\\
$(u_1-u_2)$ &-0.055 $\pm$ 0.100\\
\enddata
\end{deluxetable}

\clearpage

\figcaption{Typical spectrum of HD~209458 obtained with the STIS
instrument, showing the number of detected photons in a single
60 s integration vs wavelength.  We obtained 684 such spectra in all.}

\figcaption{Time series of the corrected intensity shown separately
for each of the 4 transits observed by HST,
with successive transits offset by -0.006 for clarity.
Note that, because the transit duration is almost 2 HST orbits,
complete temporal coverage
was not obtained for any one transit.}

\figcaption{Phased light curve for all 4 transits, assuming a planetary
orbital period of 3.52474~d.
The time series for each transit has been scaled to have the same
average intensity over the 2nd and 5th (out-of-transit) orbits.}

\figcaption{Schematic illustration of the light curve of a transiting planet.
Measurable quantities are the duration of the transit $l$, the transit
depth $d$, the ingress/egress duration $w$, and the central curvature
of the light curve $C$.
Given the orbital speed (which follows from the orbital period and
the stellar mass), these quantities determine the radii of the star
and of the planet, the orbital inclination, and the degree of limb
darkening.}

\figcaption{The $\chi^2$ surfaces for the fit of $R_p$, $R_*$, $i$,
and $(u_1+u_2)$ to the observations.
Each panel shows one 2-dimensional cut through the 5-dimensional
$\chi^2$ surface, as indicated in the Figure.
The displayed ranges in each plot are the 1$\sigma$ error intervals in
each case.
The plotted contours serve to indicate the mutual dependences of the
selected parameters, but the contour levels do not correspond to
specific significance levels.}

\figcaption{Color dependence of the transit light curve, shown as the
difference (blue-red) between the normalized fluxes in the wavelength
ranges [581.9 nm, 598.3 nm] (blue) and [598.3 nm, 637.6 nm] (red),
plotted against time from transit center.
The solid line corresponds to the best-fit model color-dependent
limb darkening law, as described by Eq. (5).}

\figcaption{Upper panel: the observed light curve folded around the
central transit time (bars) binned to 60 s sampling.
The bar height indicates the formal errors of the binned estimates.
Regions 1 and 2 are shown at larger scale in the lower panels.
Lower panels: Observed intensities (black diamonds with 1$\sigma$ error bars)
and curves for best-fit models assuming maximum ring radii of 1 $R_p$
(i.e., no rings) (solid curve), 1.8 $R_p$ (dotted curve) and
2.2 $R_p$ (dashed curve).
The latter 2 curves are inconsistent with the data.
Rings were assumed to be opaque, to lie in the planet's orbital plane,
and to extend from 1 $R_p$ to the
stated radii.}

\figcaption{Residuals around the fit of a 5-parameter transit model
(see text) to the observations, shown with error bars.
The solid lines show the light curve one would expect given the presence
of a satellite of 1.5 Earth radii, with an orbital period of 1.5 d.
Note the relatively large errors for the first transit, due to
the offset of the spectrum in the CCD subarray.}

\figcaption{The $\chi^2$ curves for fits to the central time of transit
for each of the 4 transits
(dotted = 25 April, dashed = 28/29 April, 
triple-dot dashed = 5/6 May, dot-dashed = 12/13
May).
The horizontal bar shows
the range of timings resulting from a 3-Earth-mass satellite at a
distance of 1 Hill Sphere radius.}

\clearpage

\begin{figure}
\plotone{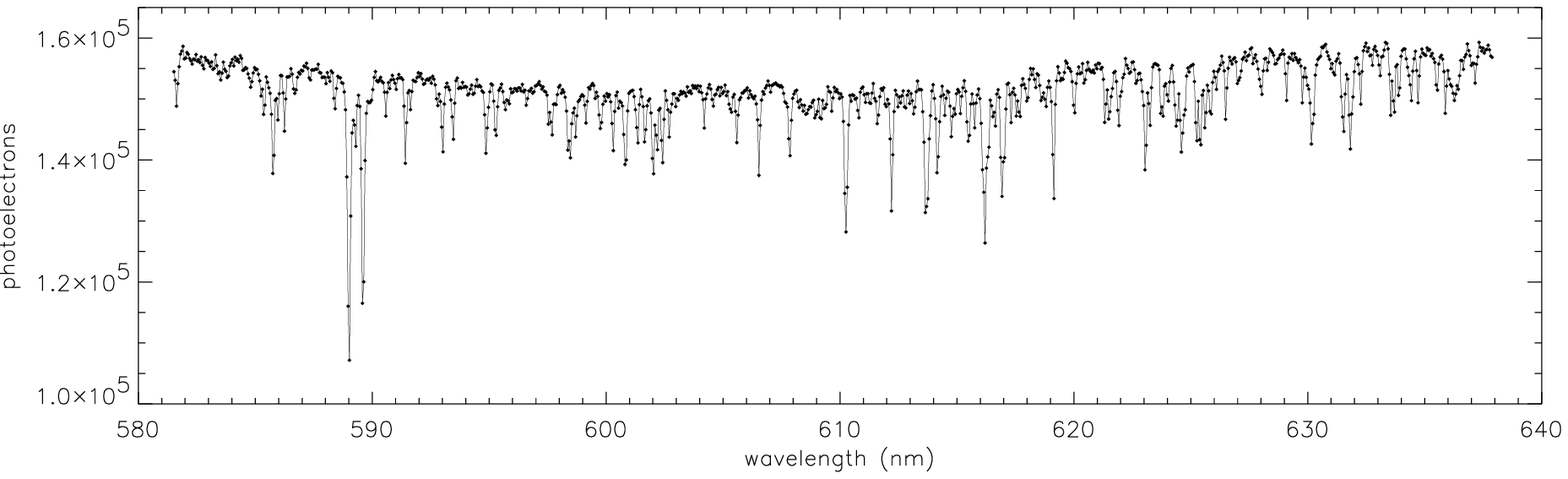}
\end{figure}

\clearpage

\begin{figure}
\plotone{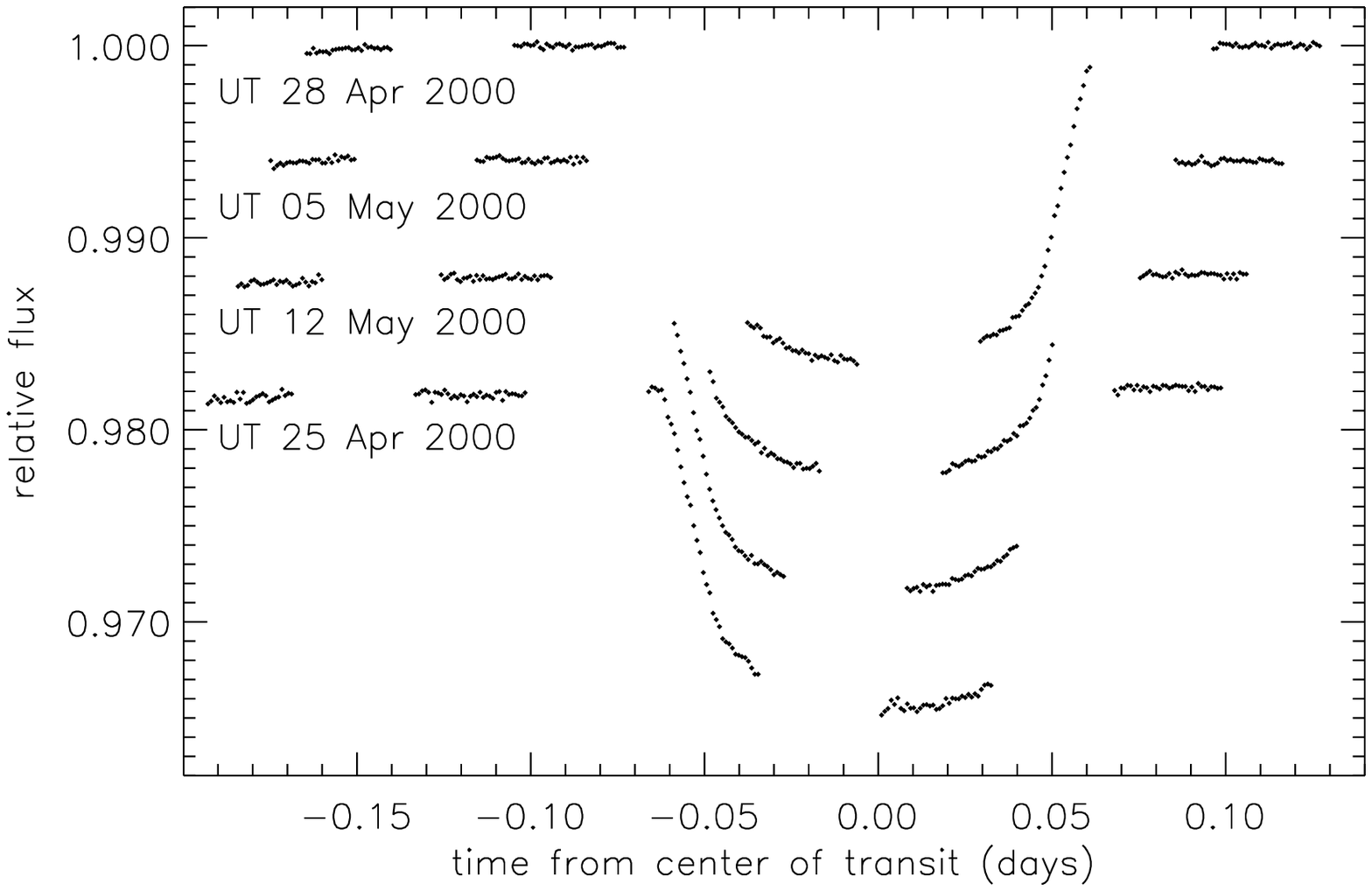}
\end{figure}

\begin{figure}
\plotone{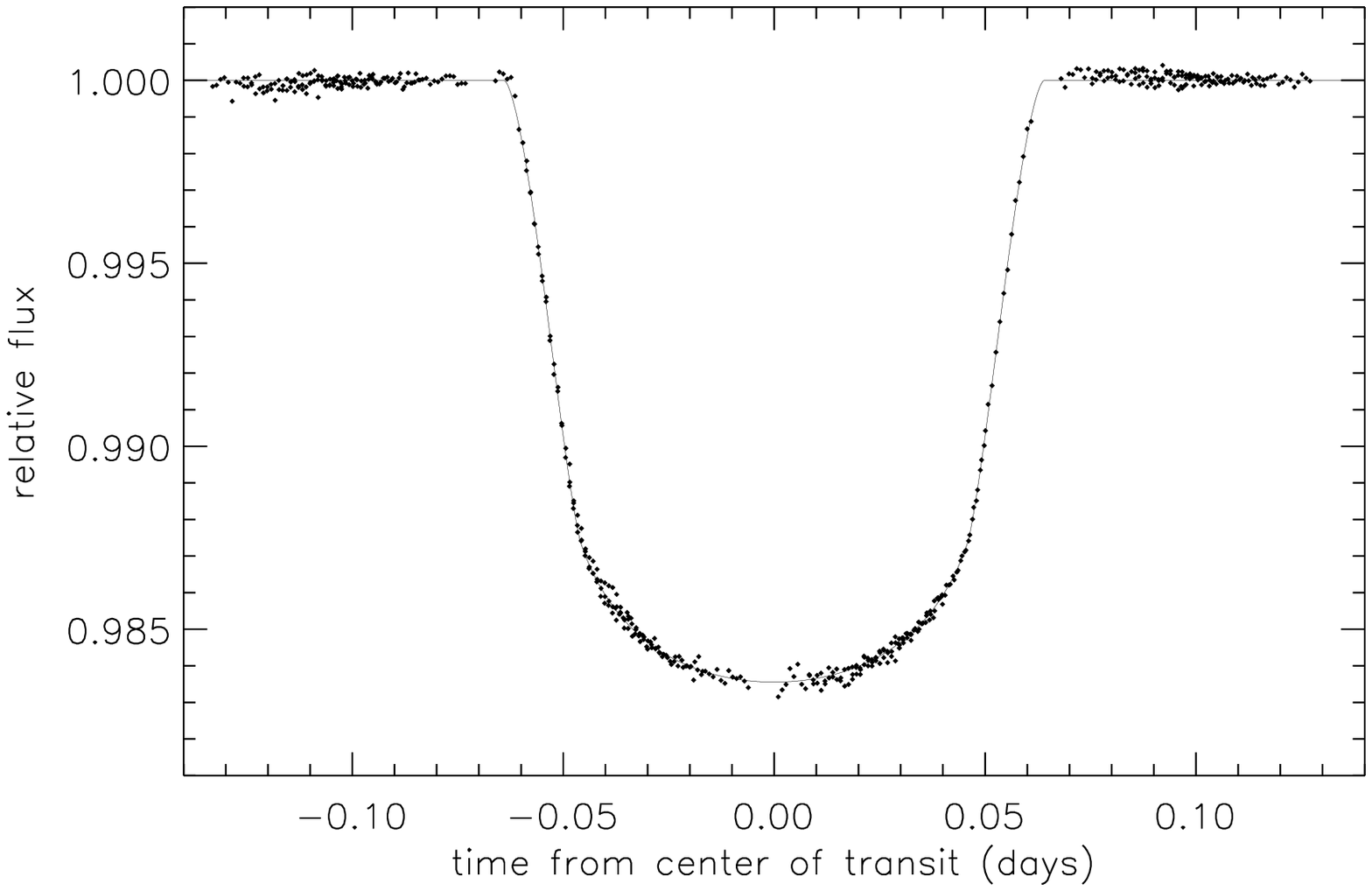}
\end{figure}

\begin{figure}
\plotone{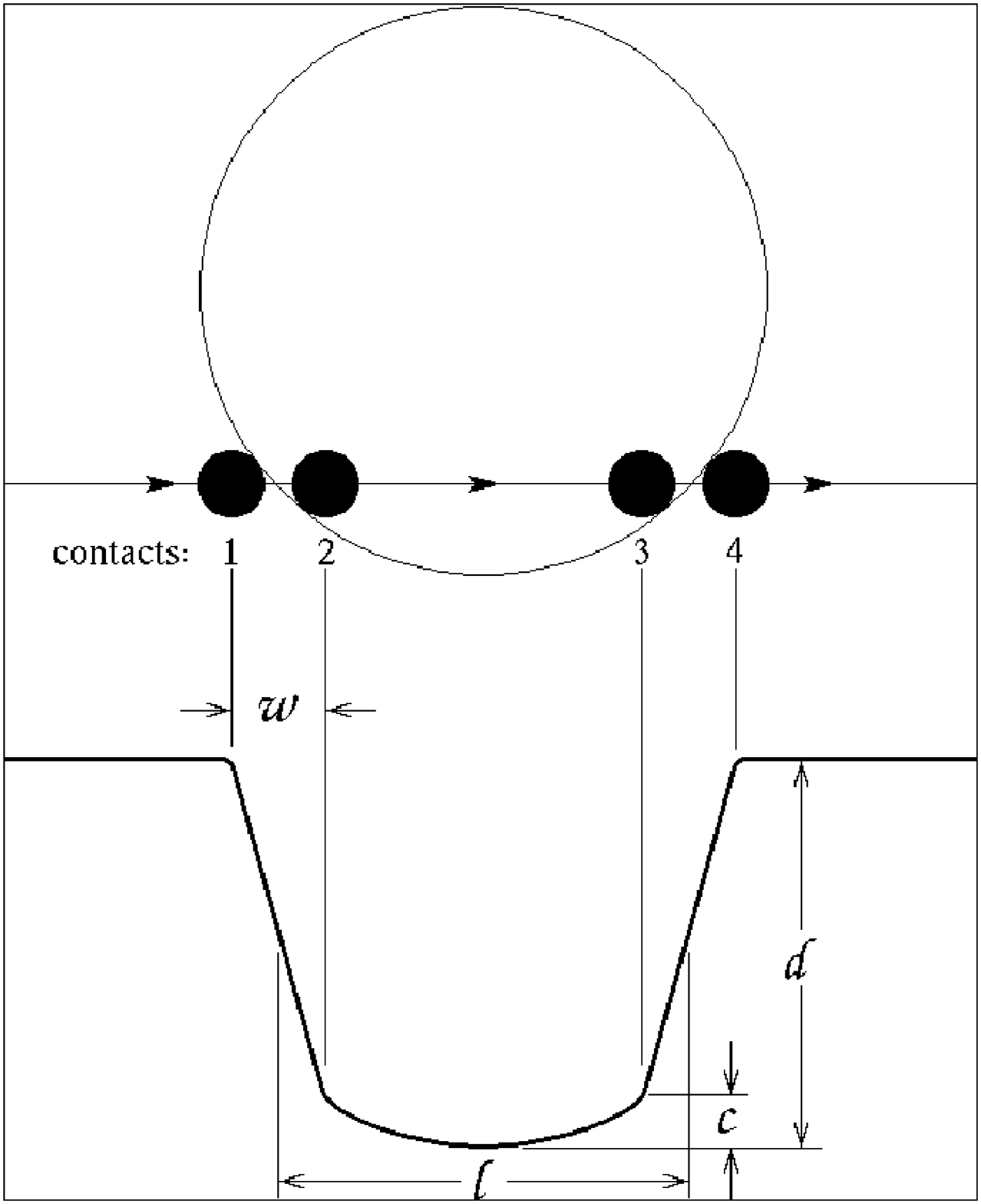}
\end{figure}

\begin{figure}
\plotone{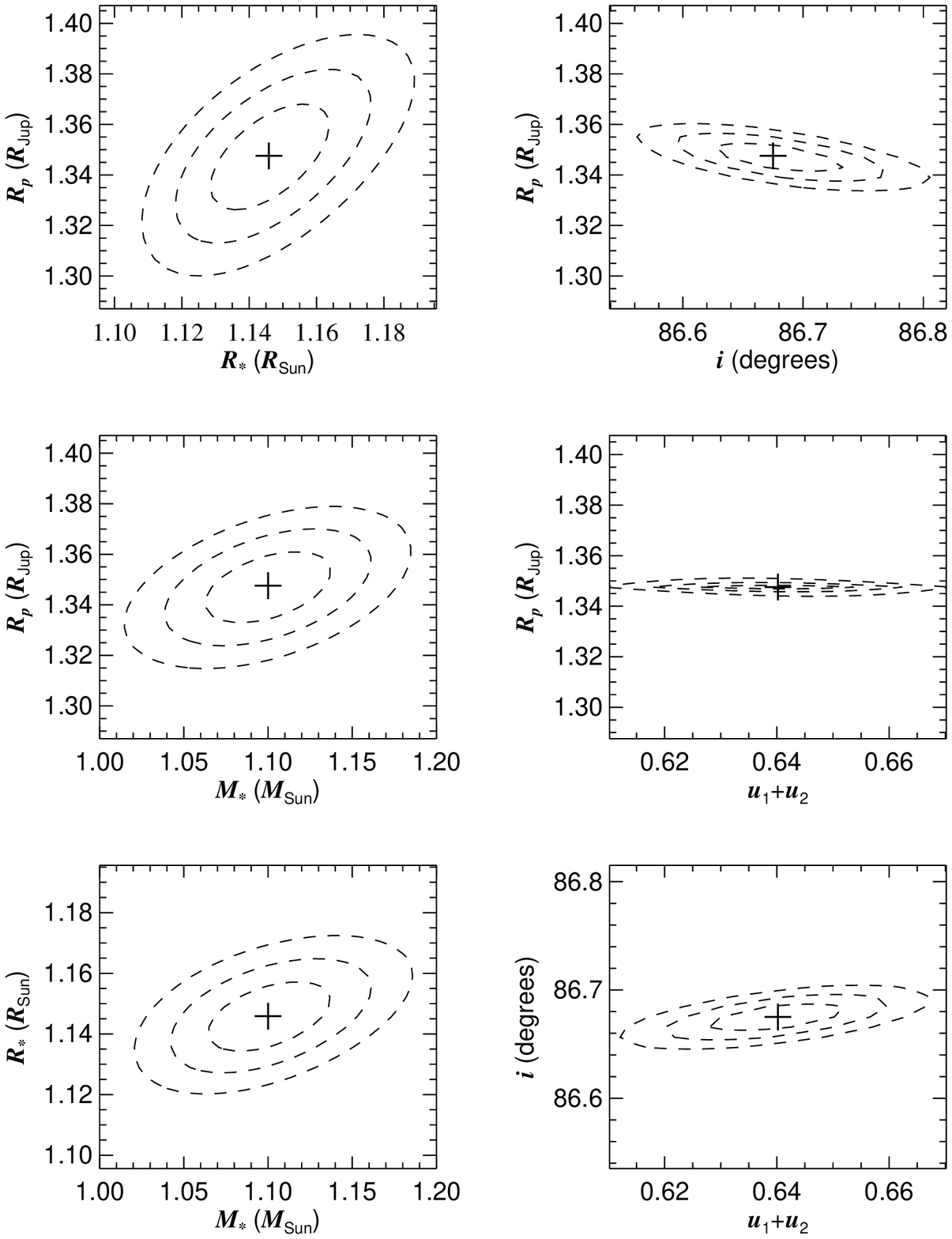}
\end{figure}

\begin{figure}
\plotone{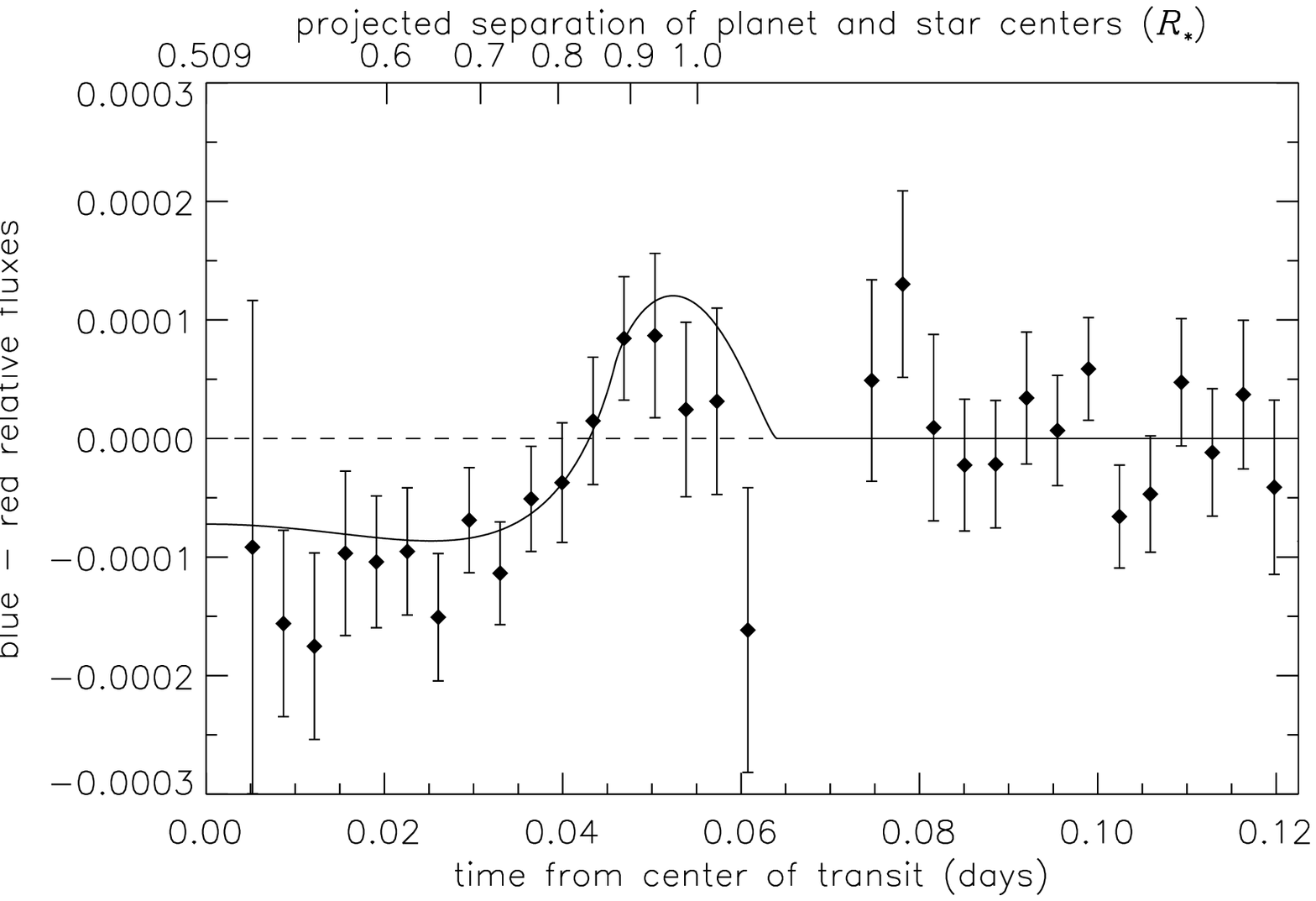}
\end{figure}

\begin{figure}
\plotone{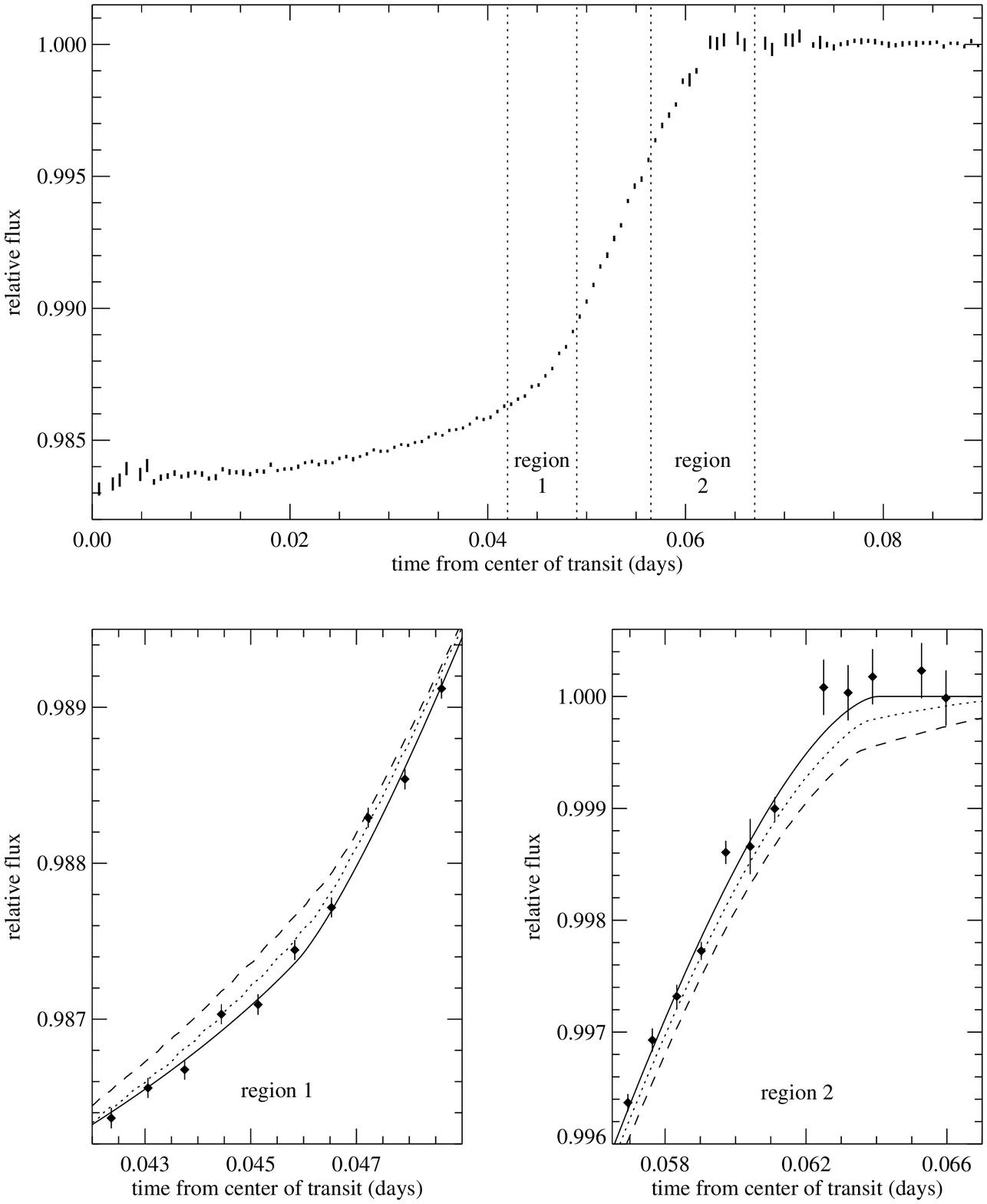}
\end{figure}

\begin{figure}
\plotone{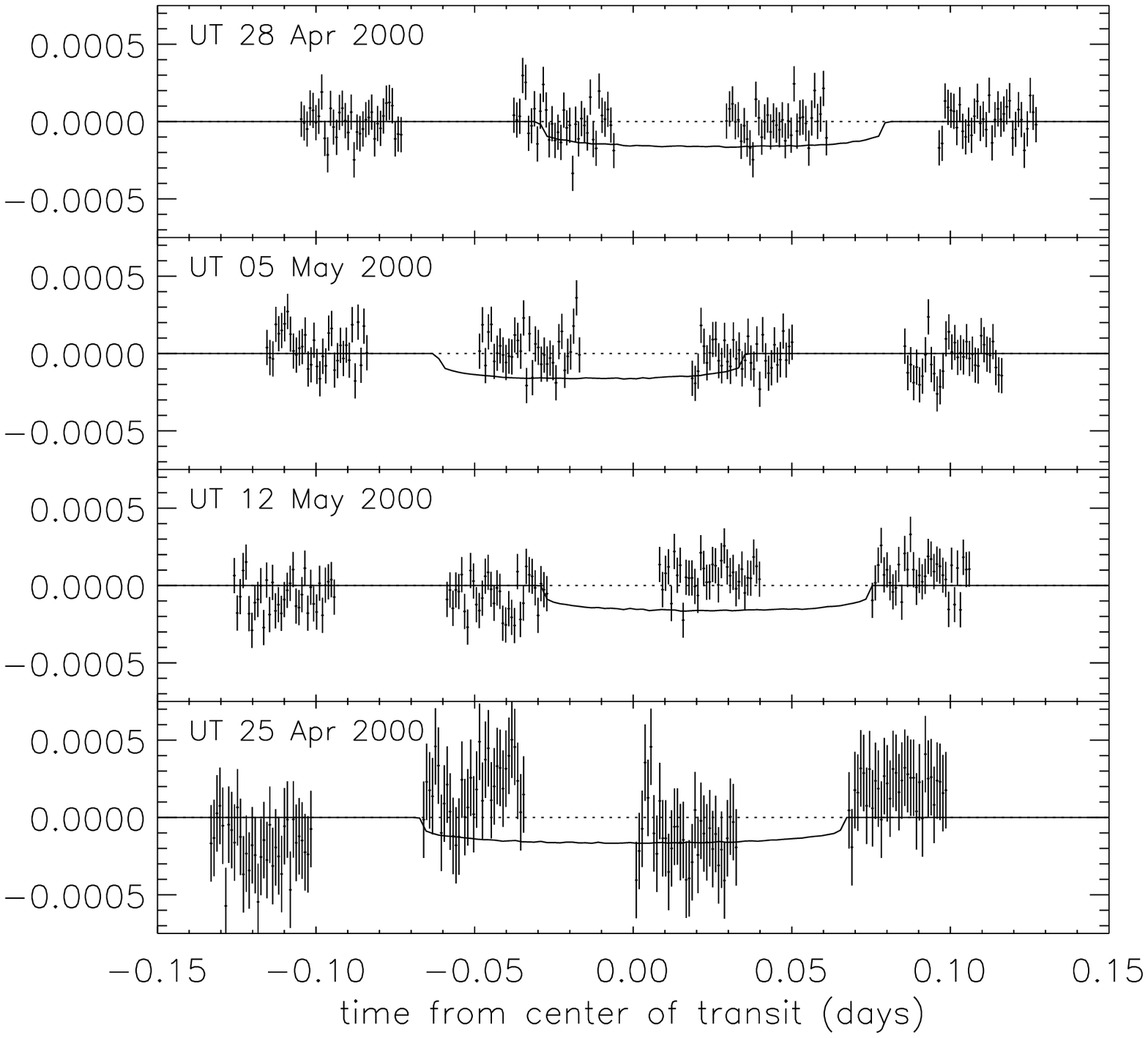}
\end{figure}

\begin{figure}
\plotone{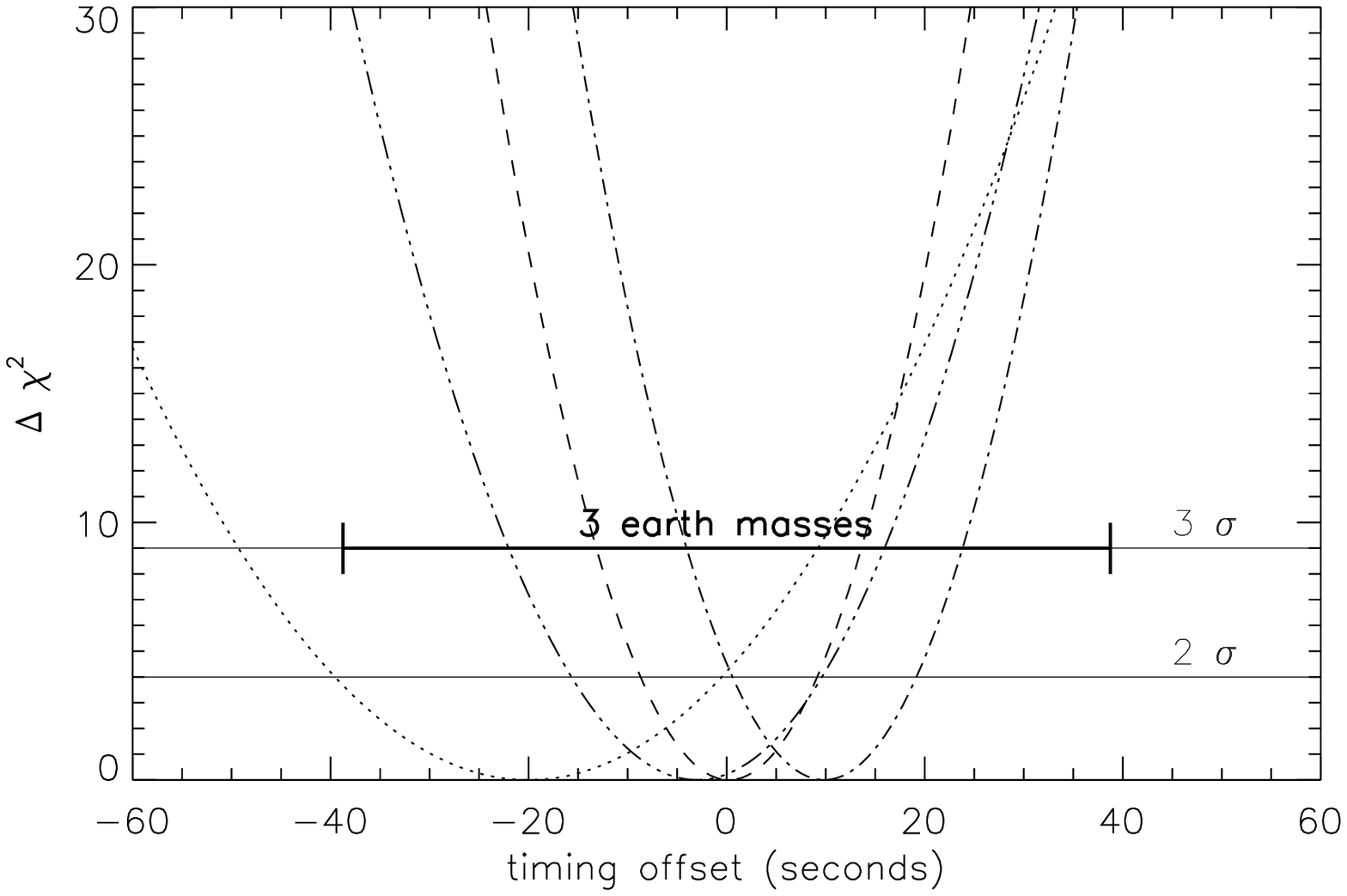}
\end{figure}


\begin{thebibliography}

\bibitem[Brown (2001)]{bro01a} Brown, T.~M. 2001, astro-ph/0101307

\bibitem[Brown et al. (2001)]{bro01} Brown, T.~M., Charbonneau, D., 
\& Gilliland, R.~L.\ 2001, \apj, in preparation

\bibitem[Burrows et al. (2000)]{bur00} Burrows, A., Guillot,
T., Hubbard, W.~B., Marley, M.~S., Saumon, D., Lunine, J.~I., \&
Sudarsky, D. 2000, \apjl, 534, L97

\bibitem[Claret \& Gimenez(1990)]{cla90} Claret, A.\ \& 
Gimenez, A.\ 1990, \aap, 230, 412

\bibitem[Collier Cameron et al. (1999)]{col99}
Collier Cameron, A., Horne, K., Penny, A., \& James, D. 
1999, \nat, 402, 751

\bibitem[Castellano et al. (2000)]{cas00} Castellano, T.,
Jenkins, J., Trilling, D.~E., Doyle, L. \& Koch, D. 2000, \apjl, 532,
L51

\bibitem[Charbonneau et al. (1999)]{cha99}
Charbonneau, D, Noyes, R.~W., Korzennik, S., Nisenson, P., Jha, S., 
Vogt, S.~S., \& Kibrick, R.~I. 1999, \apjl, 522, L145

\bibitem[Charbonneau, Brown, Latham, \& Mayor (2000)]{cha00}
Charbonneau, D., Brown, T.~M., Latham, D.~W. \& Mayor, M. 2000, \apjl,
529, L45

\bibitem[Cox (1999)]{cox99} Cox, A. N. 1999, 
Allen's Astrophysical Quantities (Springer-Verlag: New York), 355 

\bibitem[Doyle et al. (2000)]{doy00} Doyle, L.~R., et al., 
2000, \apj, 535, 338

\bibitem[Gilliland (1999)]{gil99a} Gilliland, R.~L. 1999,
STIS Instrument Science Report 99-05

\bibitem[Gilliland, Goudfrooij, \& Kimble (1999)]{gil99b} Gilliland, R.~L.,
Goudfrooij, P., \& Kimble, R.~A. 1999, \pasp, 111, 1009

\bibitem[Gilliland et al. (2000)]{gil00} Gilliland, R.~L., et al. 
2000, \apjl (in press)

\bibitem[Guillot et al. (1996)]{gui96} Guillot, T., Burrows,
A., Hubbard, W.~B., Lunine, J.~I., \& Saumon, D. 1996, \apjl, 459, L35

\bibitem[Guillot (1999)]{gui99} Guillot, T. 1999, Science,
286, 72

\bibitem[Henry, Marcy, Butler, \& Vogt (2000)]{hen00} Henry,
G.~W., Marcy, G.~W., Butler, R.~P., \& Vogt, S.~S. 2000, \apjl, 529,
L41

\bibitem[H{\o}g et al. (2000)]{hog00} H{\o}g, E., et al. 
2000, \aap, 355, L27

\bibitem[Jha et al. (2000)]{jha00} Jha, S., Charbonneau, D.,
Garnavich, P.~M., Sullivan, D.~J., Sullivan, T., Brown, T.~M., \& 
Tonry, J.~L. 2000, \apjl, 540, L45

\bibitem[Kjeldsen, Bedding, \& Christensen-Dalsgaard (2000)]
{kje00} Kjeldsen, H., Bedding, T.~R., \&
Christensen-Dalsgaard, J. 2000, in ASP Conf. Ser. 203, 
The Impact of Large-Scale Surveys on Pulsating Star Research, 
ed. L. Szabados \& D. Kurtz, (San Francisco: ASP),73

\bibitem[Koch et al. (1998)]{koc98} Koch, D.~G., Borucki, W., 
Webster, L., Dunham, E., Jenkins, J., Marriott, J., \& Reitsema, 
H.~J. 1998, \procspie, 3356, 599

\bibitem[Marley et al. (1999)]{mar99} 
Marley, M. S., Gelino, C., Stephens, D., Lunine, J. I., \& 
Freedman, R. 1999, \apj, 513, 879

\bibitem[Matthews et al. (2000)]{mat00} Matthews, J.~M., et al. 2000, 
in ASP Conf. Ser. 203: The Impact
of Large-Scale Surveys on Pulsating Star Research, 
ed. L. Szabados \& D. Kurtz, (San Francisco: ASP),74

\bibitem[Mazeh et al. (2000)]{maze00} Mazeh, T., et al. 
2000, \apjl, 532, L55

\bibitem[Michel et al. (2000)]{mic00} Michel, E., et al. 
2000, in ASP Conf. Ser. 203, The Impact of
Large-Scale Surveys on Pulsating Star Research, 
ed. L. Szabados \& D. Kurtz, (San Francisco: ASP), 69

\bibitem[Perryman et al. (1997)]{per97} Perryman, M.~A.~C., et al. 
1997, \aap, 323, L49

\bibitem[Robichon \& Arenou (2000)]{rob00} Robichon, N. \&
Arenou, F. 2000, \aap, 355, 295

\bibitem[Rosenblatt (1971)]{ros71} 
Rosenblatt, F. 1971, Icarus, 14, 71

\bibitem[Sackett (1999)]{sac99}
Sackett, P.~D. 1999, in Planets Outside the Solar System: Theory
and Observations, ed. J.-M. Mariotti \& D.~M. Alloin (NATO/ASI Ser C;
Dordrecht: Kluwer), 189

\bibitem[Sartoretti \& Schneider (1999)]{sart99} Sartoretti,
P. \& Schneider, J. 1999, \aaps, 134, 553

\bibitem[Schneider (1999)]{sch99} Schneider, J.\ 1999, C.R.\ 
Acad.\ Sci.\ Ser.\ II, 327, 621-634 (1999), 327, 621

\bibitem[Seager \& Sasselov(1998)]{sea98} Seager, S.\ \& 
Sasselov, D.\ D.\ 1998, \apjl, 502, L157 

\bibitem[Seager \& Sasselov (2000)]{sea00} Seager, S. \& 
Sasselov, D.~D. 2000, \apj, 537, 916

\bibitem[Sudarsky, Burrows \& Pinto(2000)]{sud00} Sudarsky, 
D., Burrows, A.\ \& Pinto, P.\ 2000, \apj, 538, 885 

\end{thebibliography}
\end{document}